\documentclass{article}
    \usepackage{arxiv}
    \usepackage{moreverb,url}
    \usepackage[hidelinks]{hyperref}
    \usepackage{IEEEtrantools}
    \usepackage[square,numbers]{natbib}

\usepackage[utf8]{inputenc}
\usepackage[T1]{fontenc}
\usepackage{graphicx}
\usepackage{amsmath,stmaryrd,mathtools}
\usepackage{amsthm}
\usepackage{amssymb}
\usepackage{bm}
\usepackage{bbm}
\usepackage{mathdots}
\usepackage{booktabs}
\usepackage{multicol}
\usepackage{multirow}
\usepackage{enumitem}
\usepackage{algorithm}
\usepackage{algpseudocode}
\usepackage{graphicx}
\usepackage{xspace}

    \newcommand{\resizeboxM}[3]{#3}
    \def\imwidth{100mm}
    \def\tabwidth{100mm}

\renewcommand{\v}[1]{\boldsymbol{#1}}
\def\eg{e.g.,\ }

\def\cf{\textit{cf.}\ }

\newcommand{\TN}{$\mathrm{TN}$\xspace}
\newcommand{\TP}{$\mathrm{TP}$\xspace}
\newcommand{\FN}{$\mathrm{FN}$\xspace}
\newcommand{\FP}{$\mathrm{FP}$\xspace}
\newcommand{\TNm}{\mathrm{TN}}
\newcommand{\TPm}{\mathrm{TP}}
\newcommand{\FNm}{\mathrm{FN}}
\newcommand{\FPm}{\mathrm{FP}}
\newcommand{\ut}{\mathrm{ut}}
\newcommand{\pf}{(1)\xspace}
\newcommand{\ps}{(2)\xspace}
\newcommand{\pt}{(3)\xspace}

\newcommand{\titleName}{Interpretable Detection of Partial Discharge in Power Lines with Deep Learning}

\newcommand{\abstractT}{Partial discharge (PD) is a common indication of faults in power systems, such as generators, and cables. These PD can eventually result in costly repairs and substantial power outages. PD detection traditionally relies on hand-crafted features and domain expertise to identify very specific pulses in the electrical current, and the performance declines in the presence of noise or of superposed pulses. In this paper, we propose a novel end-to-end framework based on convolutional neural networks. The framework has two contributions. First, it does not require any feature extraction and enables robust PD detection. Second, we devise the pulse activation map. It provides interpretability of the results for the domain experts with the identification of the pulses that led to the detection of the PDs. The performance is evaluated on a public dataset for the detection of damaged power lines. An ablation study demonstrates the benefits of each part of the proposed framework.}

\newcommand{\keywor}{Partial Discharges, Power distribution lines, Temporal CNN, Fault Detection.}

\newcommand{\aknow}{This work was supported by the Innosuisse grant no. 27662.1 PFES-ES.}

\newcommand{\graphAbst}{
\begin{figure*}
    \centering
   \includegraphics[width=\textwidth]{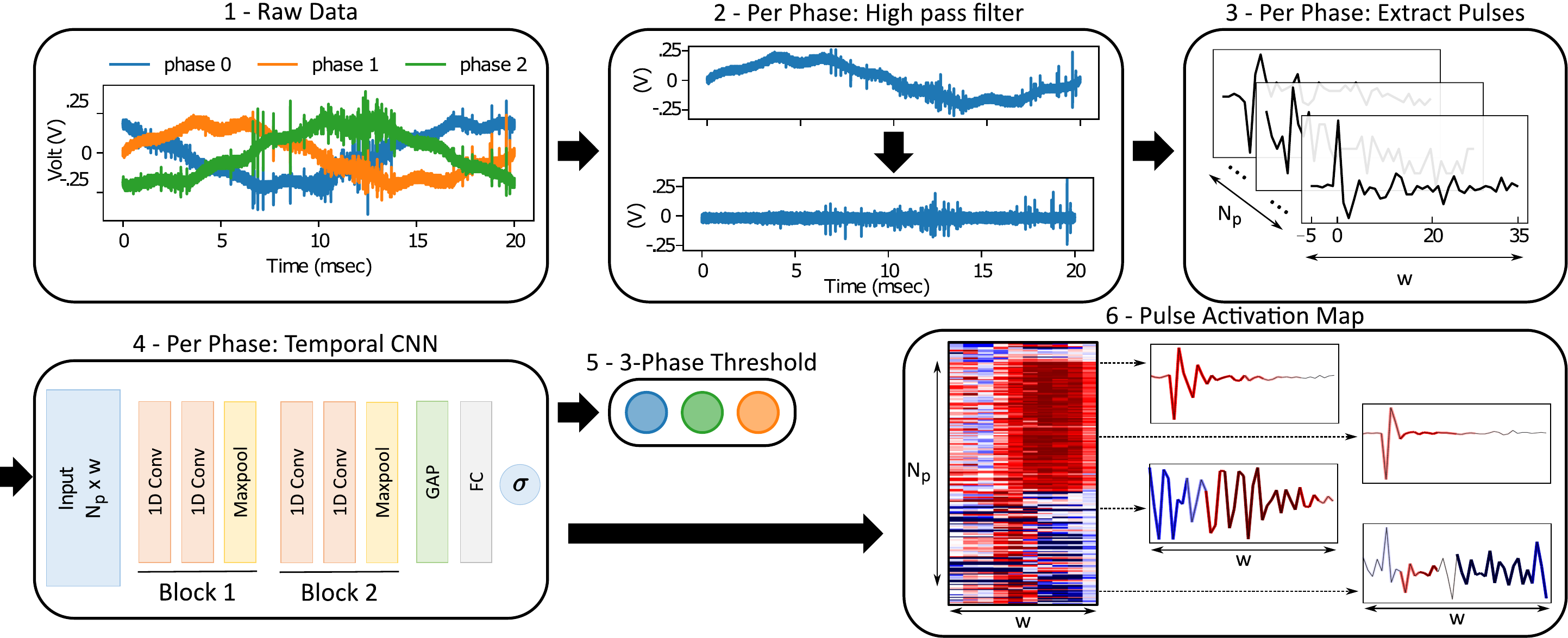}
      \caption{\textbf{Graphical Abstract:} (1) The raw data: three phases over one period. (2) For each phase, the signal is filtered to remove the utility frequency $f_{\ut}$. (3) Extracted $w$ samples of pulses identified with a maximum filter. The windows start 1/8th before the identified peak. (4) For each phase, the pulse collection is fed to a deep 1D convolution neural network. (5) The outputs of the three phases are considered together to label the powerline. (6) Thanks to the GAP Layer, the Pulse Activation Map can be displayed for further diagnostics.}
         \label{fig:abst}
\end{figure*}}

\begin{document}

    \title{\titleName}
    \author{%
    	Gabriel Michau\\
        ETH Z\"urich,\\
        Z\"urich, Switzerland\\
        \And 
        Chi-Ching Hsu\\
        ETH Z\"urich,\\
        Z\"urich, Switzerland\\
        \And 
        Olga Fink\\
        ETH Z\"urich,\\
        Z\"urich, Switzerland}
        \subtitle{Accepted in MDPI Sensors (ISSN 1424-8220)}
        \date{17th of March 2021}
        \maketitle
        \begin{abstract}
        \abstractT
        \end{abstract}

\section{Introduction}
\label{sec:intro}
\graphAbst

Many critical services in our society, such as healthcare, transportation and security require a robust, reliable and undisrupted supply of electricity. This requires on the one hand a reliable and redundant infrastructure, and on the other hand, the ability to maintain the performance of the infrastructure. As such, the ability to detect faults and to act accordingly is of critical importance for maintaining a high availability of the network~\cite{chen2017combined}. 
Many components of the power generation and distribution network can be directly monitored with specific sensors. However, it is not possible or not cost efficient for all the components and all the fault types. Therefore, for some of the components, the monitoring can only be performed indirectly through the behaviour of the electrical current. For example, insulation damages in power systems, such as generators, or defects in medium voltage cables~\cite{mashikian2006medium} can be monitored by detecting localised pulses in the electrical current, namely partial discharges (PD). 

According to the IEC 60270 international standard, Partial discharges are ``localized electrical discharges that only partially bridge the insulation between conductors''. In fact, the presence of PD can be indicative of anomalies in many electrical systems and can cause further degradation of the insulation. The high voltage discharges deteriorate the insulation materials and can have impacts on the entire system. Their detection is, therefore, of utmost importance to assess the condition of electrical components and has been a long-standing challenge ~\cite{kawaguchi1969partial}. As such, the literature is extremely vast. PD detection has been studied in many systems such as in transformers~\cite{ibrahim2012realization}, gas-insulated high-voltage switchgear~\cite{khan2019partial}, power plants~\cite{mcgreevy2017deployment}, and power lines~\cite{misak2017complex}.
The main challenge in PD detection lies in the detection of extremely short and temporally localised events: their wavelength is at the micro-second scale. It requires, therefore, extremely high frequency data (several tens of MHz). In addition, only few pulses can occur per period of the current utility frequency (usually 50 or 60Hz)~\cite{kawaguchi1969partial}. In brief, PD signs in the electrical current represent roughly 1/20\,000-th of the data. Until the very recent development of technologies able to capture and store such vast amount of data, the detection of PD patterns had to be performed online.

Among the traditional approaches, a group of approaches takes advantage of the property that for some systems, PD always occurs at the same phase in the electrical current. These approaches are also referred to as the phase-resolved partial discharge (PRPD) detection methods~\cite{hudon2005partial}. They consist in detecting pulses in the electrical current, whereby the simplest method to detect a pulse is to apply a maximum filter. Subsequently, the detected rate of occurrence (n) of pulses is plotted as a function of their voltage amplitude (Q) and of their phase value ($\phi$). It can be implemented online and experts can inspect PRPD graphs to recognise the patterns generated by the different types of PD~\cite{strachan2008knowledge}. However, to obtain meaningful occurrence rates, these methods require the aggregation of pulses over several hundreds or thousands of periods. Overall, these methods are expensive, since experts need to constantly monitor the $\phi$-Q-n diagrams. The interpretation of the diagrams also becomes difficult in the presence of noise or of superimposed pulses. 

Since all types of PD are not necessarily resolved in the phase domain, statistical approaches are also frequently used~\cite{karami2019feasibility}. They aim at characterizing the pulses with engineered features. The features become a multi-dimensional space where decision boundaries are established, for example with traditional classifiers such as random forest~\cite{misak2017complex}, support vector machines (SVM)~\cite{raymond2017classification, dong2019pattern}, or deep learning methods such as artificial neural networks (ANN)~\cite{karimi2019novel}, convolutional neural networks (CNN)~\cite{li2016partial,banno2018partial}, autoencoders~\cite{ganjun2018partial}, and recurrent neural networks (long short-term memory networks LSTM)~\cite{nguyen2018recurrent, dong2020partial, qu2020fault}. For a more exhaustive overview, the reader is referred to the literature reviews in~\cite{wu2015overview, barrios2019partial}. These methods are time consuming and expensive due to the difficulty of extracting and engineering the relevant features. It requires years of domain expertise for a profound understanding of the system. Additionally, the methods suffer from degradation performances when pulses are superimposed. 

To address the aforementioned challenges, we propose to take advantage of the recent advances in Deep Learning (DL) applied to time series (TS) anomaly detection. For example, DL has already been successfully applied to identify cardiac abnormalities in electrocardiography (ECG) data~\cite{acharya2017deep}. In particular, convolutional neural networks (CNN) have recently demonstrated very good performances for TS classification~\cite{Susto2018,fawaz2019deep}, forecasting~\cite{Zeng2019Research} and anomaly detection~\cite{Habeeb2019}. In fact, temporal convolutions are able to learn meaningful filters in the time domain, adjusted to the signature of the analysed events, and are, thus, often compared to learnable spectral features~\cite{kuo2016understanding}.

In this paper, we propose an end-to-end learning framework for partial discharge detection in time series. The framework comprises two parts: 1) the automatic PD detection without any feature engineering and 2) the subsequent extraction of pulse activation maps that provide the domain experts a possibility to interpret the results. 
The difficulty met by previous research in the detection of partial discharges lies in the discrimination between PD and non-PD related pulses. Therefore, we propose here to extract a collection of pulses for each period of the utility frequency. We train a temporal convolution neural network with the binary information of whether PD are present in the original time series or not. 
Since all pulses of the collection are processed with the same temporal filters, a well-performing model should be able learn the PD pulse signatures. 
Learning these signatures is, in fact, particularly valuable for the experts to potentially  distinguish between different types of PD. Therefore, we design our framework to provide both competitive results in terms of partial discharge detection, and a visualisation of the neural network processing of the inputs through the pulse activation maps. 
These activation maps provide interpretability and explainability of the results and allow the experts to diagnose for each time series, which pulses and which part of the pulses where dominant in the final score of the network. It gives an extremely fine interpretation of the network decision. 
We demonstrate the performance of the proposed approach by achieving rank 4 on private leaderboard of the Kaggle VSB Power Line Fault Detection competition. The aim is to identify damaged power lines from the observed PD in the electrical voltage~\cite{Kaggledata, mashikian2006medium}. Furthermore, we also demonstrate the added value of each part of the proposed framework with an ablation study. 

The paper is organized as follows: Section~\ref{sec:method} presents each step of the framework, including the pre-processing, the network architecture and the pulse activation map devising. Section~\ref{sec:exp} details the experiments we performed to quantify the results presented in Section~\ref{sec:results}. The final results are discussed in Section~\ref{sec:dis}.

\section{Methodology}
\label{sec:method}

An overview of the proposed framework is presented in Figure~\ref{fig:abst}. For each measurement (1), each phase is handled independently. Low frequencies, in particular the utility frequency, are filtered first (2). Pulses are identified and ranked with a simple maximum filter, and extracted into a pulse collection (3). Each phase is estimated independently by the same neural network (4). The final decision on the power line takes the results from the three phases into account and applies a global threshold (5). Last, a pulse activation map (6) is computed to understand which part of the input led to the networks classification results.

\subsection{Pre-processing}
The main assumption of the proposed framework relies on the inherent definition of a PD signature as a pulse in the electric current. Thus, inspired by the PRPD analysis, in the very first step, we identify and extract the pulses with a simple maximum filter. This requires first the removal of the low frequencies.

\textbf{Data filtering - Fig~\ref{fig:abst}~(2).}
PD are due to insulation failures and typically occur at very specific voltage changes. Their frequency is much higher than the utility frequency $f_{\ut}$. Thus, we first implement a high-pass filter with cutoff frequency $f_c > f_{\ut}$. In this work, we apply a Butterworth filter of order 5~\cite{butterworth1930theory}. However, other filters could be explored. An example before and after the filtering of an electric current recording over one period is shown in Figure  \ref{fig:abst}~(2), where the sampling frequency $f_s = 40MHz$, $f_c = 20kHz$ and $f_{\ut} = 50Hz$. Low frequencies such as the underlying sine-wave are eliminated after the high-pass filter, while the high frequency pulses remain unchanged.

\textbf{Pulse extraction - Fig~\ref{fig:abst}~(3).} Since the partial discharge signature is inherently a pulse in the electric potential, we propose as a second step to extract a large collection of pulses from the recordings that will be used as inputs to the neural network (NN). The goal of the NN is to learn to recognise if there is a PD pulse signature within the input collection. Due to the nature of partial discharge, we can expect some periodicity in the occurrence of the pulses with respect to the utility frequency. We create therefore, for each period of the utility frequency, a 2D array where each row represents a single pulse. The columns represent the time dimension. The number of columns corresponds to the number of timestamps $w$ collected for each pulse.

The pulses are identified with a maximum filter on the absolute value of the electric potential. The filter extracts the local maximal values which are further apart than a given window size. For simplicity purposes, we set this window to $w$. 
We extract in the filtered data, the $w$ timestamps around each of the $N_p$ largest local maxima, with an offset of $1/8\cdot w$. That is, if the $i$-th local maximum is localised at timestamp $t_i$, we extract the interval $[t_i - \frac{w}{8}, t_i+\frac{7w}{8}]$.
The collection of pulses is, therefore, a 2D array of shape $N_p \times w$. Figure~\ref{fig:abst}~(3) illustrates the pulse extraction and the resulting collection for one period and phase of the utility frequency.

$N_p$ and $w$ are hyperparameters of the proposed approach. Expert domain knowledge could help to identify relevant values for these hyperparameters. The selected values would primarily depend on the noise level for the selection of $N_p$ (some noise pulses may dominate the PD pulses), and on the expected frequency of the pulses for the selection of $w$.  Yet, if this knowledge is not available, as in our case, these hyperparameters can be selected based on cross-validation. 

\subsection{Temporal Convolutional Neural Network - Fig~1~(4)}
\label{ssec:TCNN}

\textbf{Convolutional neural network (CNN).} 
Inspired by the recent advances in computer vision and more recently also in time series classification tasks, using CNN architectures, we propose to apply a CNN for this PD detection task. For the applied architecture, we propose a deep learning neural network architecture with a similar structure to VGG neural networks~\cite{simonyan2014very}.
Yet, the architecture requires several adaptations to the high-frequency time series data as used in this case study. 
Unlike in images, where the neighbourhood of a pixel has a clear meaning in both the $X$ and the $Y$ dimensions, in the extracted pulses, only the temporal dimension contains a physically meaningful neighbourhood relationship. The pulses have been ordered by decreasing amplitudes and not by their relationships in the signal. We, therefore, apply 1D convolutions instead of 2D kernels. This also means that the temporal filters are applied identically to each pulse, performing operations similar to spectral analysis.

\textbf{Global Average Pooling (GAP).} A limitation when using CNN is that the convolutional layers preserve the dimensionality of their inputs. Therefore, since predictions are usually vectors (with one element per class), it is necessary to flatten the latent space in order to transition toward fully connected layers. A consequence is an explosion in the number of model parameters, often leading to over-fitting effects and harming the generalization ability of the network. We propose, therefore, to use the Global Average Pooling (GAP) as a structural regularizer~\cite{he2016deep, lin2013network}.  GAP takes the average over the feature maps channel-wise and thus shrinks the size of the last latent space before its vectorisation.

\textbf{Proposed CNN Architecture.} 
The proposed network architecture takes advantage of 1D CNN and the GAP layer. It contains 2 blocks comprising 2 successive convolutional layers and a max pooling layer. The 2 blocks are followed by a GAP layer, a fully connected (FC) layer and a single neuron layer for binary classification.

\subsection{Pulse Activation Maps (PAM) - Fig~1~(6)}
To provide more interpretability and more insights on the network decision to classify a collection of pulses as belonging to a damaged line or not, we propose to devise the Class Activation Maps (CAM) of our network~\cite{zhou2016learning}. 
Following the methodology in~\cite{zhou2016learning}, we devise in this section the pulse activation map (PAM) for the proposed network architecture. The PAM enables to interpret which part of the pulse has contributed most to the classification result (in this case, PD or non-PD). There are two differences to the original contribution. First, our network has a binary output. We devise a single PAM per input, instead of a per-class activation map. Second, our network contains a fully connected layer between the GAP and the output. In the following we demonstrate that the CAM (or here, the PAM) can still be computed in such cases, as long as the activation functions used by the intermediate fully connected layers are piece-wise linear. 

Since the first two blocks of our network use 1-D convolution filters, the latent space $M$ after the last block can be written as %
$M=\left(\v{m}_{plj}\right)_{p\in \llbracket1, N_p\rrbracket, l\in \llbracket 1, N_l\rrbracket, j\in \llbracket 1, N_f\rrbracket}$%
, where $N_p$ is the number of pulses used as input to the network, $N_l$ is the resulting size after the successive max-pooling (here $N_l = \lfloor\frac{w}{4}\rfloor$) , and $N_f$ is the number of filters of the last convolution. In the following, we denote the size of $M$ as $N_m = N_p\cdot N_l$

The $j$-th neuron of the GAP layer performs the operation:
\begin{equation}
\forall j \in \llbracket 1,N_f \rrbracket,\quad \mathrm{GAP}^j = \frac{1}{N_m}\sum_{p,l}^{N_P\times N_l} \v{m}_{plj}
\end{equation}

The GAP layer is connected to a fully connected layer of size $N_{FC}$ with the weights and bias %
$(\v{w}^{FC}_{ij}, \v{b}^{FC}_i)_{i\in \llbracket1, N_{FC}\rrbracket, j\in \llbracket 1, N_f\rrbracket}$%
, we have for the $i$-th neuron, before activation:

\begin{IEEEeqnarray*}{ll}
\IEEEeqnarraymulticol{2}{l}{\forall i \in \llbracket 1,N_{FC}\rrbracket,}\\
\mathrm{FC}^{i} & = \v{b}^{FC}_i + \sum_{j=1}^{N_f} \v{w}^{FC}_{ij} \cdot \frac{1}{N_m} \sum_{p,l}^{N_p\times N_l} \v{m}_{plj}\\
 & = \frac{1}{N_m} \sum_{p,l}^{N_p\times N_l} \left[ \v{b}^{FC}_i + \sum_{j=1}^{N_f} \v{w}^{FC}_{ij} \cdot \v{m}_{plj}\right] \yesnumber
\end{IEEEeqnarray*}

After the piece-wise linear activation function, the $i$-th activated neuron $\mathrm{AFC}^i$ is given by:
\begin{IEEEeqnarray*}{ll} \label{eq:afc}
\mathrm{AFC}^{i} & = \mathrm{ReLU} \left( \mathrm{FC}^{i} \right)\\
& = \delta_i \cdot \mathrm{FC}^{i}\yesnumber\\
\end{IEEEeqnarray*}
where 
\begin{equation}
\delta_i = \left\lbrace \begin{array}{ll}
		1 & \mbox{if } \mathrm{FC}^{i}>0\\
		0 & \mbox{otherwise}
		\end{array}\right.
\end{equation}
Note that the definition of $\delta$ can be generalised to any piece-wise linear activation function. Under this assumption, it is also worth noticing that Equation~\ref{eq:afc} can be generalised to as many successive fully-connected layers as required.

Last, the activated neurons are combined into a last dense layer of output size $1$. Denoting its weights and bias by %
$(\v{w}^{Out}_i,b^{Out})_{i\in \llbracket1, N_{FC}\rrbracket}$%
. The final score before the sigmoid is obtained as:

\begin{multline}
\label{eq:score}
\mathrm{Score}= b^{Out} + \sum_{i=1}^{N_{FC}} \v{w}^{Out}_i \cdot \mathrm{AFC}^{i}\\
\resizeboxM{0.9\columnwidth}{!}{=\frac{1}{N_m} \sum_{p,l}^{N_p\times N_l} \Biggl[b^{Out} + \sum_{i=1}^{N_{FC}} \v{w}^{Out}_i \delta_i \left(\v{b}^{FC}_i + \sum_{j=1}^{N_f} \v{w}^{FC}_{ij} \v{m}_{plj}\right)\Biggr]}
\end{multline}

Finally, the pulse activation map $\mathrm{PAM}$ for each input is a collection of $N_p$ vectors of size $N_l$ and is defined as:
\begin{multline}
\label{eq:pam}
\forall (p,l) \in \llbracket 1,N_p\rrbracket \times\llbracket 1,N_l\rrbracket,\\
\resizeboxM{0.9\columnwidth}{!}{\mathrm{PAM}(p,l) = b^{Out} + \sum_{i=1}^{N_{FC}} \v{w}^{Out}_i \delta_i \left(\v{b}^{FC}_i + \sum_{j=1}^{N_f} \v{w}^{FC}_{ij} \v{m}_{plj}\right)}
\end{multline}

Since the decision of the network is taken after applying the sigmoid operation to the $\mathrm{Score}$ value (Eq.~\eqref{eq:score}), we can interpret the PAM as follows: A map which average is positive corresponds to a score above 0.5 after the sigmoid operation, and is thus originating from pulses containing PD. On the contrary, a map which average is negative corresponds to a non-PD pulse. The activation maps can be used by domain experts to further evaluate the pulses and possibly to distinguish between different types of PD. 

\section{Experiments}
\label{sec:exp}

\subsection{Datasets}
To demonstrate the benefit of our approach, we apply the proposed methodology on the VSB dataset, generated and released by the Technical University of Ostrava~\cite{Kaggledata}. The goal of the case study is to detect damaged three-phase, medium voltage overhead power lines~\cite{mashikian2006medium}. According to the dataset description, damaged power lines can be identified through the observed PD patterns~\cite{Kaggledata}.  To this end, the electric voltage is recorded over one period of the grid utility frequency, 50 Hz, for the three phases simultaneously. The sampling frequency is $40MHz$ such that each recording contains 800\,000 values. An example signal is shown in Figure~\ref{fig:abst}~(1).

The VSB dataset contains two sets of measurements.
The training set contains 8,712 samples with 3 labels: the measurement ID, the phase, and whether the power line insulation was damaged at the time of recording. Damaged power lines should contain PD, however no additional information is provided on the PD types, shapes or location.
In this set, 575 samples are labeled as damaged power lines. 

The second set contains 20,037 samples with two labels: the measurement ID and the phase. No ground truth is provided with respect to the presence of PD. However, the predictions of the health state can be evaluated online through the Matthew Correlation Coefficient (MCC).

To the best of our knowledge, no other published study outside of the competition leaderboard reported results on the second test dataset. In \cite{dong2019pattern,dong2020partial,qu2020fault}, the reported results are computed on a subset of the labelled dataset. In \cite{dong2019pattern}, results are reported on the full training set and might therefore be over-fitted. In \cite{dong2020partial}, results are reported on an artificially augmented set containing 807 non-PD signals and 935 signals with PD, which might also therefore suffer over-fitting. We report anyway their results in Table~\ref{tab:exp_val} where we recompute the value of the metrics they would achieve on our set, assuming constant sensitivity and specificity of their model. This can not be done for ~\cite{qu2020fault} since the numbers of tested samples with and without PD are not reported.

\subsection{Network Architecture and Training}

The proposed neural network architecture as presented in Figure~\ref{fig:abst}~(4) comprises two convolutional blocks: The first block contains two temporal convolutional layers with $16$ kernels of size $15$. The second block has two temporal convolutional layers with $8$ kernels of size $10$. Each block is followed by a 1D temporal max-pooling layer with kernel size $2$. Therefore, the input size is $N_p \times w \times 1$, the latent space size after the first block is $N_p \times \frac{w}{2} \times 16$, the latent space size after the second block is $N_p \times \frac{w}{4} \times 8$, and the latent space size after the GAP layer is $8$. In particular, we have $N_l = \frac{w}{4}=10$ in Eq.~\eqref{eq:pam}.
The fully connected layer after the GAP layer is of size $32$. All layers but the last output layer use \textit{ReLU} as activation function. The hyper-parameters of this architecture (number of blocks, kernel number and size) were inferred from a grid search with a 5-fold stratified cross validation.

We implemented the network with Keras and TensorFlow. For the training, we used the \textit{ADAM}~\cite{kingma2014adam} optimiser with constant learning rate of $1e-3$, $\beta_1 = 0.9$, and $\beta_2 = 0.999$. We used the binary cross entropy loss:
\begin{equation}
\label{eq:binary_ce}
    L = - (y log(p) + (1-y)log(1-p)),
\end{equation}
where $y$ is the ground truth and $p$ is the network output.

\subsection{Threshold setting}
The problem at hand is a binary classification problem. The output is, therefore, designed as a single neuron output layer, activated with a sigmoid function such that the output is continuous between 0 and 1. The traditional baseline consists in using a threshold at value 0.5 on the network output value. Compared to the baseline approach, we propose to explore two modifications: first, the inference of an optimised threshold $th_v$ based on a validation set; second, the consideration of the three phases as a single indicator of the power line health. We propose, thus, to compare four different post-processing approaches of the network output:
\begin{enumerate}[label={(\roman*)}]
    \item \label{itm:trad}\textbf{Baseline:} Round the output (threshold $th=0.5$).
    \item \label{itm:1=3}\textbf{`1=3'-Phase Classification:} Round the prediction of the three phases. If one phase is estimated as damaged, the whole power line (the three phases) is considered as being damaged.
    \item \label{itm:thv}\textbf{1-Phase Optimized Threshold:} Infer the threshold $th_v$ with cross-validation, and consider each phase independently.
    \item \label{itm:3thv}\textbf{Proposed 3-Phase Global threshold:} Using the threshold $th_v$ devised in~\ref{itm:thv}, apply the 3-Phase Global threshold, $G_{th_v}=3\cdot th_v $, to the sum of the output values for the three phases.
\end{enumerate}
Please note that in~\ref{itm:3thv}, the direct inference of the $G_{th_v}$ with cross-validation instead of using $3\cdot th_v$ may have enhanced the performance. This is, however, not possible since in the training set, some samples do not have all the three phases recorded as damaged. 
Using 5-fold stratified cross validation to maximise the MCC (defined below in Section~\ref{ssec:metrics}), we define $th_v = 0.28$.

\subsection{Evaluation Metrics}
\label{ssec:metrics}
As part of the competition, a tool is provided to evaluate the test set online. It provides an evaluation of the results with the Matthews Correlation Coefficient ($\mathrm{MCC}$)~\cite{Matthews1975Comparison}:  
\begin{equation}
\label{eq:mcc}
\resizeboxM{\columnwidth}{!}{
    \mathrm{MCC} = \frac{(\TPm\cdot\TNm)-(\FPm\cdot\FNm)}{\sqrt{(\TPm+\FPm)(\TPm+\FNm)(\TNm+\FPm)(\TNm+\FNm)}}}
\end{equation}
where \TP is the number of true positives, \TN is the number of true negatives, \FP is the number of false positives, and \FN is the number of false negatives. This metric varies from -1 to 1, where 1 indicates an optimal solution, 0 a solution no better than a random guess and -1 a total disagreement with the ground truth.

To enrich our evaluation of the network performance, we propose to use in addition three common evaluation metrics for binary classification problems, namely, \textit{Accuracy}, \textit{Precision}, and \textit{Recall}, as defined in Equation~\eqref{eq:acc},\eqref{eq:pre}, and \eqref{eq:recall}.

\begin{equation}
\label{eq:acc}
    Accuracy = \frac{\TPm+\TNm}{N}
\end{equation}
where $N= \TPm + \TNm + \FPm + \FNm$ is the total number of samples.

\begin{equation}
\label{eq:pre}
    Precision = \frac{\TPm}{\TPm+\FPm}
\end{equation}

\begin{equation}
\label{eq:recall}
    Recall = \frac{\TPm}{\TPm+\FNm}
\end{equation}

If the \textit{Accuracy} can give a false sense of performance of the model in a strongly imbalanced dataset (a naive model always classifying to the main class would have high \textit{Accuracy}), its derivatives with respect to \TP and \TN are identical and constant. Changes in \textit{Accuracy} are, therefore, easier to interpret than in other metrics.

\section{Results}
\label{sec:results}

\subsection{Results on the Test Dataset}
\label{ssec:tets}
\begin{table}
\centering
\caption{Results on the Test Dataset (evaluated online), for our four proposed decision rules and for $N_p $ set to $200$ and $150$.}
\begin{tabular}{lcc|cc}
\toprule
\multirow{2}{*}{Settings} & \multicolumn{4}{c}{Kaggle VSB test data}  \\
 & \multicolumn{2}{c|}{$N_p = 150$, $w = 40$} & \multicolumn{2}{c}{$N_p = 200$, $w = 40$}\\ \midrule
 & MCC  private & Rank*   & MCC  private & Rank*\\ \midrule

\ref{itm:trad} \textit{Baseline}   & 0.580 & 1054 & 0.629 & 272\\
\ref{itm:1=3} \textit{`1=3'}      & 0.656 & 118 & \textbf{0.704} & \textbf{4}\\
\ref{itm:thv} \textit{1-Phase Opt.} & 0.631 & 245  & 0.658 & 65\\
\ref{itm:3thv} \textit{3-Phase (our)} & \textbf{0.665} & \textbf{46} & \textbf{0.704} & \textbf{4}   \\
\bottomrule
\multicolumn{5}{c}{* Ranking on the private leaderboard of Kaggle}\\
\multicolumn{5}{c}{(First place: MCC=0.719)}
\end{tabular}
    
\label{tab:exp}
\end{table}

We evaluated the predictions of the model on the test dataset, using the tool provided online by the hosting platform.
Table~\ref{tab:exp} summarises the MCC and the rank reported, for the 4 proposed decision rules (\cf Section~\ref{sec:exp}, \ref{itm:trad}-\ref{itm:3thv}), for $N_p=200$ and for $N_p=150$. Remind that $N_p$ represents the number of extracted pulses used as input. 

The obtained results show that, first, our proposed approach provides state-of-the-art results. With settings \ref{itm:1=3} and \ref{itm:3thv} it achieves \textbf{rank 4} on the leaderbord of the competition, and this is to our knowledge the best results achieved so far with a pure deep-learning approach, without careful feature engineering. For comparison, the first place achieved an MCC of $0.719$ (+2\%), using a LightGBM model on carefully designed features using manually tuned threshold.

Second, the results demonstrate that considering $N_p = 200$ pulses per $50Hz$ period leads to optimal results. This illustrates that the partial discharge pulses are not always the most dominant pulses in the signal. Since the pulses are selected by amplitude, selecting more pulses increases the likelihood of capturing PD pulses. We also evaluated the performance for a larger number of pulses. However, the performance started to decrease, this may be explained either by the noise, or by the fact that increasing the input size also increases the number of parameters of the model and may lead to over-fitting. The training dataset comprising around $8\,000$ samples is rather small from a deep-learning perspective.

\subsection{Ablation Study}

In addition to evaluating the performance of the proposed framework on the test dataset, we performed an ablation study to evaluate the impact of the different composing elements of the proposed network in more detail. 
Since the test dataset does not contain any ground truth information, we cannot perform a more detailed analysis. Thus, we decided to split the training dataset into a new training dataset of size 6\,972 (6\,538 non-damaged and 434 damaged power-line samples), and a new test set of size 1\,740, (1\,599/141).

Table~\ref{tab:exp_val} summarizes the results of the ablation study. The evaluated experiments comprise the proposed model with the four decision rules (as discussed above) in part \pf, and the ablated versions of this model. First, without the use of the Global Average Pooling layer (`w/o $GAP$') in part \ps, second, without the last fully connected layer between the GAP layer and the output (`w/o $FC$') in part \pt. We only present the results for $N_p = 200$ and $w=40$ since this was the best performing parameter on the cross-validation.

\begin{table}
\centering
\caption{Ablation study and impact of the decision rules on the proposed model with $N_p = 200$ and $w=40$.}
\begin{tabular}{llcccc}
\toprule
& Methods  & MCC & Acc. & Precision  & Recall \\
\midrule
\multirow{4}{*}{\pf} & \ref{itm:trad} \textit{Baseline}     & 0.710  & 0.959 & \textbf{0.772} & 0.695\\
&\ref{itm:1=3} \textit{`1=3'} & 0.802 & 0.963 & 0.687 & \textbf{0.979} \\
&\ref{itm:thv} \textit{1-Phase Opt.} & 0.727  & 0.953   & 0.669 & 0.844   \\
&\ref{itm:3thv} \textit{3-Phase (our)} & \textbf{0.817} & \textbf{0.967} & 0.726 & 0.957 \\

\midrule
\multirow{4}{*}{\ps} &w/o $GAP$ - \ref{itm:trad} & 0.583  &0.944 & 0.712  & 0.525  \\
&w/o $GAP$ - \ref{itm:1=3} & 0.722 &0.952 & 0.656  & 0.851  \\
&w/o $GAP$ - \ref{itm:thv} & 0.598 & 0.945 & 0.709 & 0.553 \\
&w/o $GAP$ - \ref{itm:3thv} & 0.616 & 0.955 & 0.683 & 0.610  \\
\midrule
\multirow{4}{*}{\pt} &w/o $FC$ - \ref{itm:trad} & 0.643   & 0.952 & 0.784  & 0.567    \\
&w/o $FC$ - \ref{itm:1=3} & 0.751 & 0.959 & 0.702 & 0.851 \\
&w/o $FC$ - \ref{itm:thv} & 0.701 & 0.954 & 0.700  & 0.745  \\
&w/o $FC$ - \ref{itm:3thv} & 0.775 & 0.966 &0.772 & 0.816 \\
\midrule
 & (STL + SVM)$^{\dagger}$ & 0.779$^\triangleright$ & 0.963$^\triangleright$ & 0.73$^\triangleright$/0.68$^*$ & 0.88$^*$\\
 & (LSTM)$^{\dagger\hspace{-0.5mm}\dagger}$ & 0.344$^\triangleright$ & 0.765$^\triangleright$ & 0.23$^\triangleright$/0.79$^*$ & 0.81$^*$\\
\bottomrule
\multicolumn{6}{p{\tabwidth}}{$^{\dagger}$ \footnotesize{Reported results on training set \cite{dong2019pattern}}}\\
\multicolumn{6}{p{\tabwidth}}{$^{\dagger\hspace{-0.5mm}\dagger}$ \footnotesize{Reported results on samples drawn in the artificially augmented training set (split: non-PD: 807 / PD: 935) \cite{dong2020partial}}}\\
\multicolumn{6}{p{\tabwidth}}{$^*$ \footnotesize{Metrics as reported.}}\\
\multicolumn{6}{p{\tabwidth}}{$^\triangleright$ \footnotesize{Metrics recomputed on our own data split assuming constant sensitivity and specificity of the model.}}\\
\end{tabular}
\label{tab:exp_val}
\end{table}
\section{Discussion}
\label{sec:dis}

\subsection{Results Analysis}

\textbf{The Global Threshold provides the best results:} The results both of the ablation study and of the evaluation on the test dataset (Tables~\ref{tab:exp} and ~\ref{tab:exp_val}) show that setting \ref{itm:3thv} always outperforms others both in terms of \textit{MCC}  and in terms of  \textit{Accuracy}. The only exception is the network without GAP layer. Since this architecture performs worst overall, the comparison of the settings may not be very insightful in that case.

\textbf{3-phase classification improves the performance:}
In general, the results demonstrate that considering the three phases together (settings \ref{itm:1=3} versus \ref{itm:trad} and \ref{itm:3thv} versus \ref{itm:thv}) is always improving the overall performance (\textit{MCC} and \textit{Accuracy}), independently of the other parameters. This also makes sense from an application perspective since the essential information for the operator is to know whether a power line is damaged or not. Furthermore, some types of damage may impact all phases.

\textbf{Optimising the threshold favors the \textit{Recall} over the \textit{Precision}:}
Looking at the \textit{MCC} in all setups, it appears that decreasing the detection threshold to an optimised threshold improves the results (comparing the settings \ref{itm:thv} versus \ref{itm:trad} and \ref{itm:3thv} versus \ref{itm:1=3}). This is in fact due to the non-linearity of MCC, which favors the detection of the smallest class (the true positives, \TP) strongly over the true negatives (\TN). Comparing settings \ref{itm:thv} versus \ref{itm:trad} in the part \pf of Table~\ref{tab:exp_val} illustrates the strong impact of the revised threshold on the \textit{Recall}, while the \textit{Precision} is harmed. From an application perspective, it seems indeed more important to detect true positives, even if it increases the number of false alarms. This is especially true in our case since a follow-up expert confirmation of positive cases is simplified thanks to the Pulse Activation Map presented in the next section. On the contrary, unnoticed damaged power-lines can lead to cascading damages and stronger consequences on the power distribution system.

\textbf{Single-Phase Detection also provides a good performance:} 
An additional interesting observation is that even though considering the detection results of 3 phases jointly provides the best performance, the proposed model is still able to provide good detection performance when the phases are considered independently (settings \ref{itm:trad} and \ref{itm:thv} in Table~\ref{tab:exp_val}). This is especially true for setting \ref{itm:thv}. This confirms that the proposed framework can also be used on single phase measurements if all three phases are not available simultaneously.

\textbf{The GAP layer is essential for the performance of the proposed architecture:} The ablation study also confirms the significance of the GAP layer. Irrespective of the settings, the \textit{MCC} always decreases when the GAP layer is removed from the architecture (part \ps of Table \ref{tab:exp_val}). The biggest impact of the GAP layer appears to be on the \textit{Recall}, that is the ability of the model to correctly identify true positives. 
As discussed in Section~\ref{ssec:TCNN}, GAP improves the generalisation ability of the networks by decreasing the number of model parameters. For the network used here with $N_p = 200$ and $w=40$, using the GAP layer instead of a flatten layer reduces the number of parameters from $518\,113$ to $6\,369$. The resulting model is less prone to over-fitting, which may explain its better performance. 
In addition, the GAP layer allows us to extract the Pulse Activation Map as described in Section~\ref{ssec:pam}.

\textbf{The last FC layer improves the \textit{Recall} significantly:} Contrary to the architectures commonly encountered in the literature, adding a FC layer between the GAP and the output improves the results in our case study. This can be inferred by comparing part \pf and \pt of Table~\ref{tab:exp_val}. The absence of the FC layer seems to harm the discriminative capacity of the network since the \textit{Precision} increases slightly while the \textit{Recall} decreases strongly. In fact, the model identifies more true negatives \TN but far less \TP. The \textit{Accuracy} is in fact on par with the best models. However, the \textit{MCC} strongly decreases. As already discussed above, from an application perspective, we believe that the \textit{Recall} is more important than the \textit{Precision}. This is in line with the competition objective of \textit{MCC} maximisation.

\textbf{Comparison with the state-of-the-art:}  The proposed network outperforms state of the art methodologies found in the literature  on both the \textit{Recall} and the \textit{MCC} and would be considered as superior according to the competition objective.

\subsection{Note on the MCC for imbalanced datatset}
\begin{figure}
  \centering
   \includegraphics[width=\imwidth]{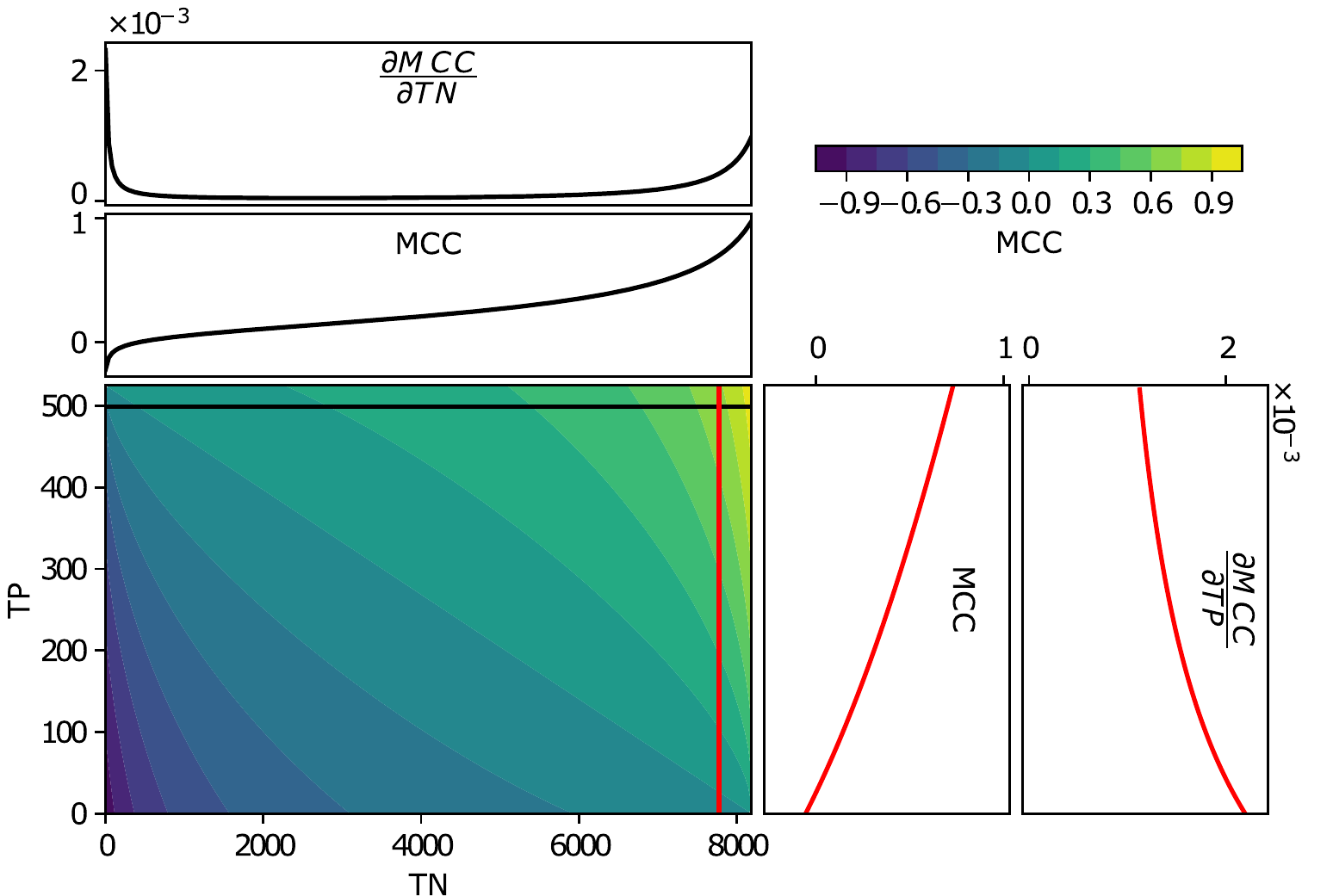}
      \caption{\textbf{MCC:} Colormap of the MCC. From colormap to top, \TP$=95\%$, MCC and its derivative as a function of \TN.  From colormap to right, \TN$=95\%$, MCC and its derivative as a function of \TP.}
         \label{fig:MCC}
\end{figure}

The results in Table~\ref{tab:exp_val} indicate that higher \textit{MCC} are always linked to higher \textit{Recall} values, but not always to higher \textit{Precision} values. This is in fact explained by the higher partial derivative of the \textit{MCC} with respect to the true detection of the smallest class (here, true positives) compared to the detection of the largest class (here, true negatives). This is illustrated in Figure~\ref{fig:MCC}, where the \textit{MCC} is represented using a colormap as the function of both \TP and \TN. On each side,  a slice of the map $MCC$ is represented when either the \TP-rate or the \TN-rate are respectively set to $95\%$. In Figure~\ref{fig:MCC}, the top sub-plot and rightmost sub-plot represent the corresponding partial derivatives.  Excluding the region of low \TN, the partial derivative of the MCC with respect to the \TP is always much higher compared to the  partial derivative with respect to the \TN.
Since only the \textit{MCC} is reported for the test dataset by the online tool, it gives little insights on the impact of the different experimental setups on the \TP and \TN values. Changes in \TP would dominate similar changes in \TN. As discussed previously, the use of this metric is justified from an application perspective, since it favors \textit{Recall} over \textit{Precision}.

\subsection{Pulse Activation Maps (PAM)}
\label{ssec:pam}

\begin{figure}
\centering
\includegraphics[width=\imwidth]{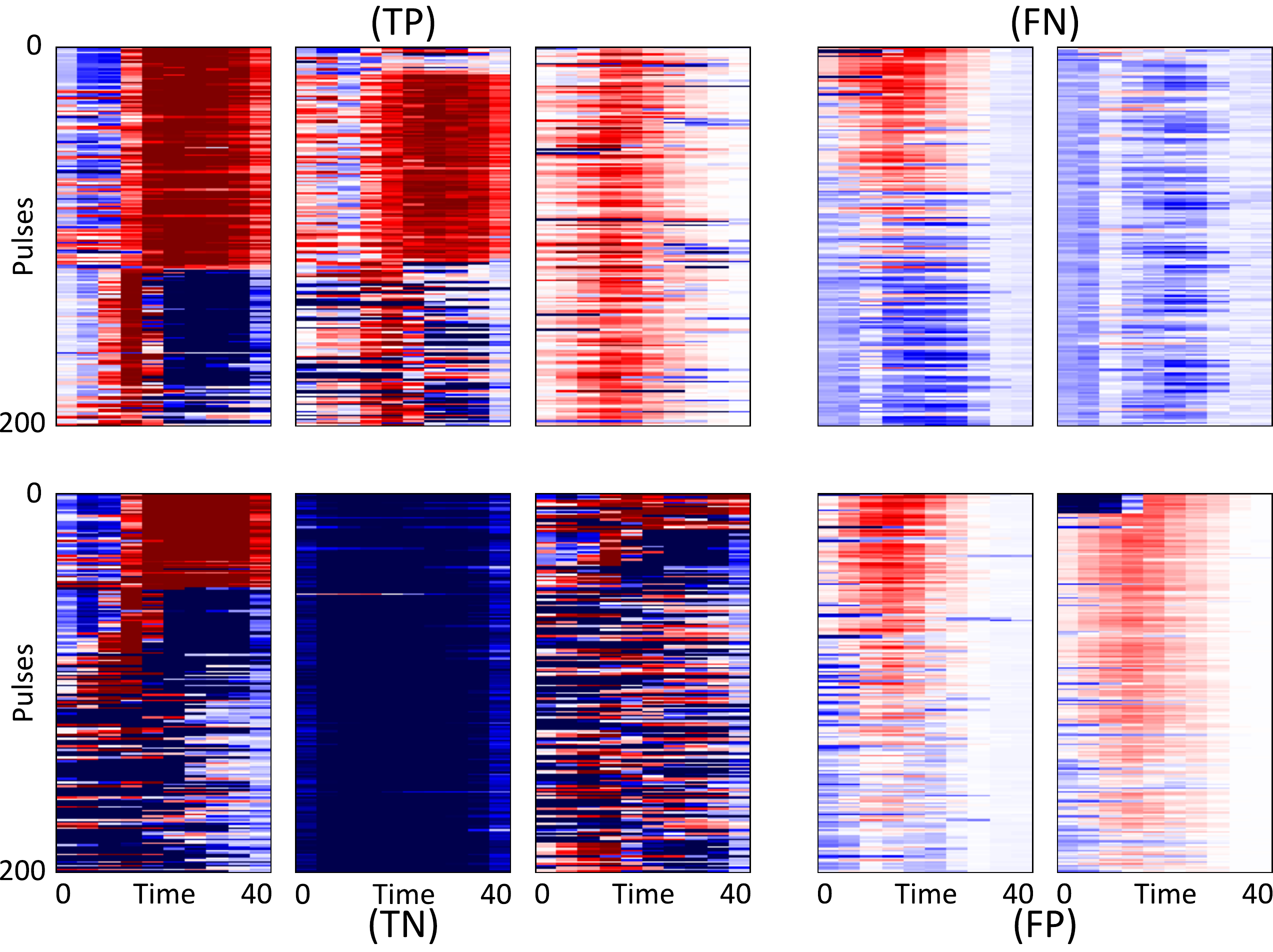}
\caption{\textbf{PAM:} For \TP (3 upper left), \FN (2 upper right), \TN (3 lower left) and \FP (2 lower right). The color mapping is symmetric around 0 (white), from dark-blue (strongly negative) to dark-red (strongly positive). Note the the actual values matter little since the decision of the network only depends on whether the map is positive or negative in average.}
\label{fig:pam}
\end{figure}

\begin{figure}
\centering
\includegraphics[width=\imwidth]{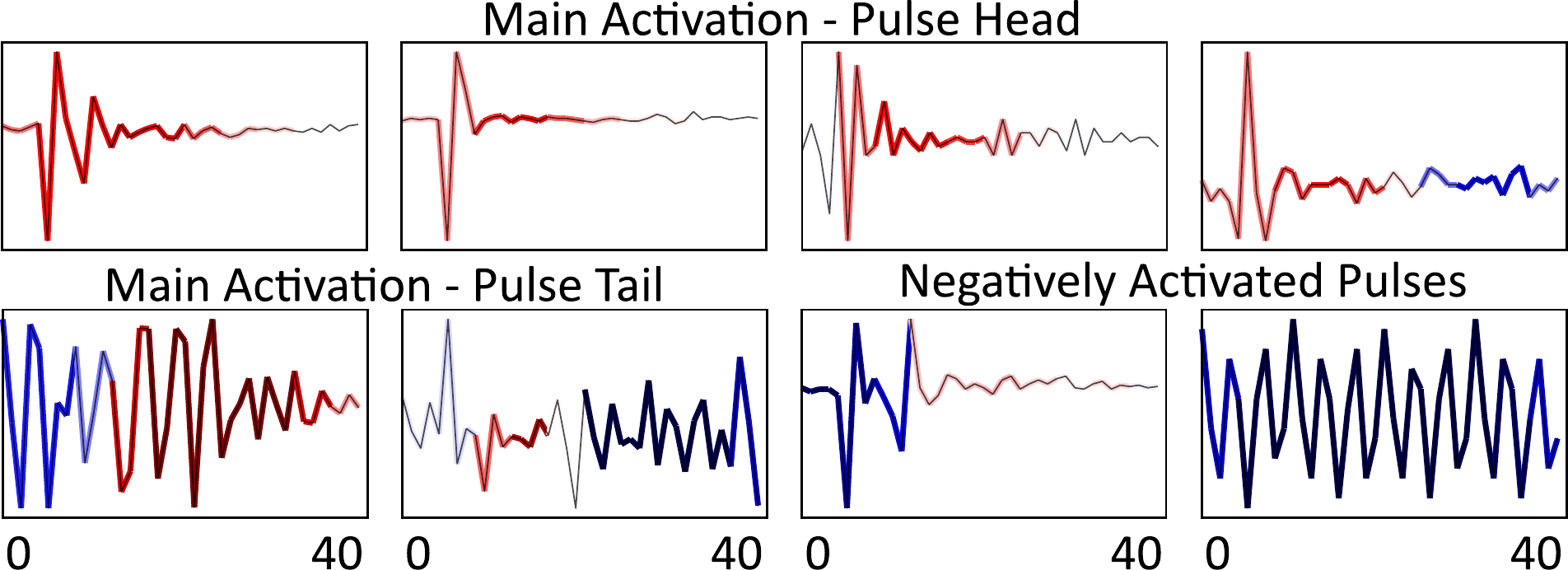}
\caption{\textbf{Pulse Activation:} Some typical pulses with their activation (same color-map as in Fig.~\ref{fig:pam}). Upper row: pulses positively activated. Lower row left, the tail is rather considered as distinctive by the network. Lower row, right, strongly negatively activated pulses (The last one appears to be noise). The $Y$-axes are independent.}
\label{fig:pam-p}
\end{figure}

To provide more insights on the decision of the network and enable interpretability of the obtained results, we propose to use the PAM as per Equation~\eqref{eq:pam} as a tool for a deeper analysis of the pulses for domain experts. Plots of some representative sample maps are demonstrated in Fig.~\ref{fig:pam}. 

Potentially, the most valuable information for the domain experts is the evaluation of the activation of the pulses, as demonstrated in Fig.~\ref{fig:pam-p}, providing an information which parts of the pulses have contributed to the respective decision of the algorithm (either positive or negative). 

Interestingly, two types of patterns appear to influence the decision (\cf Fig.~\ref{fig:pam}): Samples for which the activation has higher amplitude for the pulse tails, (\eg first \TP and first \TN), and samples for which the pulse itself is the deciding factor (last \TP, and the two displayed \FP). These different types of pulses are presented in Figure~\ref{fig:pam-p}. Remarkably, it appears that the decision to label one power line as damaged or not depends as much on the presence of positively activated parts as on the presence of pulses that are strongly deactivated (negative parts of the map). For example, the difference between the first \TP and the first \TN maps appears to rely on the number of pulses whose tails are positively activated. Also, for the last \FP, the largest pulses (the one at the top of the map) were identified as not being an indicator of a damaged power line. However, this is compensated on average by the rest of the map which is weakly positively activated. One could infer that the line may have some PD. However, those PD are not strong enough for the line to be considered as damaged. The first plotted \FN in the figure can lead to a similar interpretation. The network did find some PD pulses in the signal (upper part is mostly red). However, they were apparently not sufficiently strong to compensate for the second part of the map which is weakly deactivated (light blue).

Finally, Figure~\ref{fig:pam-p} first row, presents pulses that are only positively activated. They could, therefore, be interpreted as representative of some typical PD pulses. An input from domain experts could provide additional insights and evaluate the match of the learnt patterns to the expert intuition on the type of the PD. This would guide additionally the interpretation of the obtained results.

\section{Conclusion}
\label{sec:sum}

In this paper, we proposed a new framework for the detection of damaged power lines. The proposed approach offers several improvements with respect to traditional power-line diagnostics. 
First, the proposed framework does not require any feature engineering and is able to handle raw measurements with extremely little pre-processing. 
Second, it provides competitive detection results at the power-line level, but also at the phase level. The proposed  approach is robust and can detect damages in power lines from a single period of utility frequency. It provides a significant speed up compared to the more traditional PRPD approaches that require firstly, the processing of several hundreds of periods, and secondly, an expert analysis of the diagrams.

In addition, we proposed to extract the Pulse Activation Maps to improve the interpretability and to gain understanding on which part of the electrical signals are learned by the network as being a signature of a damaged power line. PAM can be used by the domain experts to gain more insights in the decisions of the proposed neural networks and to perform the diagnostics. PAM provide the information on which pulses and  which part of the pulses dominated the decision of the neural network and allows to verify the network's decision. 

It can be pointed out that one limit of the task we tackled here is the relatively small size of the training dataset (from a deep learning perspective). Even though very competitive results were obtained, we believe our approach can showcase its full potential when more and more data will be available. Training the framework with more data will allow for a more precise tuning of the hyper-parameters. Furthermore, if samples were identified per power-line, timestamped and collected over a long time period (which was not the case in the considered case study), the monitoring of the PAM evolution over time would be a very promising follow-up research. We could expect that, as a power-line damage increases, the PAM would become more and more positively activated, and such monitoring would have a potentially large benefit for the utility operators.

\section*{Acknowledgements}
\aknow

    \bibliographystyle{abbrvnat}
    \bibliography{bibliography}

\end{document}